\documentstyle[epsf,floats,twocolumn,prb,aps]{revtex}
\newcommand{\bS}{\mbox{\boldmath $S$}}
\newcommand{\bn}{\mbox{\boldmath $n$}}
\newcommand{\bm}{\mbox{\boldmath $m$}}
\newcommand{\bl}{\mbox{\boldmath $l$}}
\newcommand{\bA}{\mbox{\boldmath $A$}}

\newcommand{\bB}{\mbox{\boldmath $B$}}

\newcommand{\cH}{{\cal H}}
\newcommand{\cU}{{\cal U}}
\newcommand{\cN}{{\cal N}}

\newcommand{\stimes}{\!\times\!}
\newcommand{\scdot}{\!\cdot\!}
\newcommand{\bu}{\mbox{\boldmath $u$}}
\newcommand{\bv}{\mbox{\boldmath $v$}}
\begin{document}
\draft
\preprint{}
\title{Alternating-Spin Ladders}
\author{ Takahiro Fukui\cite{Email}} 
\address{Institute of Advanced Energy, Kyoto University,
Uji, Kyoto 611, Japan}
\author{Norio Kawakami}
\address{Department of Applied Physics,
Osaka University, Suita, Osaka 565, Japan} 
\date{May 7, 1997: Revised, August 18, 1997}
\maketitle
%----------------------------------------------------------------------
%                              Abstract
%----------------------------------------------------------------------
\begin{abstract}
We investigate a two-leg spin ladder system composed  
of alternating-spin chains with two-different
kind of spins. The fixed point properties are discussed 
by using spin-wave analysis and non-linear sigma
model techniques. The model contains various massive phases,
reflecting the interplay 
between  the bond-alternation and the spin-alternation.
\end{abstract}
%----------------------------------------------------------------------
%                           Introduction
%----------------------------------------------------------------------
\section{Introduction}

Since the seminal work of Haldane\cite{Hal},
quantum spin chains have been extensively studied as
one of the simplest but the most typical quantum many-body systems.
The fixed-point properties 
of such systems depend on whether the spin is integer or
half-integer, 
as shown by the use of the non-linear
sigma model (NLSM) with a topological term\cite{Hal,Rev}.
The massive phase of the integer-spin chains, called Haldane phase,
has been understood as valence-bond-solid (VBS) states
proposed by Affleck et al.\cite{AKLT}.
Quantum phase transitions caused by the bond-alternation have 
been predicted \cite{AffHal} and have been confirmed 
numerically \cite{KatTan}, which in fact fits in with the VBS picture.
Current interest has been spread to wider classes of spin chains,
stimulated by experimental realization of a 
variety of spin systems.
A typical example is the spin ladder system\cite{RICE,NTAFT,FNST},
owing to the discovery of the
high-temperature superconductivity, and another 
is the alternating-spin 
chain composed of two kind of spins\cite{AltExp,AltThe,AltExa}.

In this paper, we propose and investigate a novel spin model,
 stimulated by recent 
interests mentioned-above, i.e., 
a two-leg spin ladder model 
composed of alternating-spin
chains with some kind of bond-alternations.
This type of spin systems are expected to 
be synthesized experimentally in future, and thus could provide an  
interesting example in quantum spin systems.
What is particularly interesting in this model is that 
{\it one can explicitly see the 
interplay between the bond-alternation and the spin-alternation,
both of which affect the quantum phase transitions.}
We first study the model by the spin-wave analysis, and 
then by mapping it to the NLSM 
we demonstrate that it is not only the bond-alternation 
but also the  spin-alternation 
which gives rise to a rich structure for the 
phase diagram.

%------------------------------------------------------------------------
%                      II. Model
%------------------------------------------------------------------------
\section{The model}

%------------------------------------------------------------------------
%   FIGURE of model
%------------------------------------------------------------------------
\begin{figure}[htb]
\epsfxsize=5cm %%% 6 is suitable for 2 column
\centerline{\epsfbox{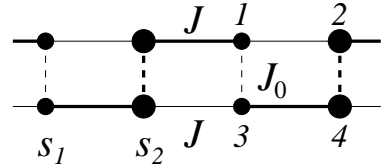}} 
\vspace{0.2cm}
\caption{Schematic illustration of the model. }
\label{f:SheIll}%------------------------------------------------------
\end{figure}
The model we investigate (see Fig.\ref{f:SheIll}) 
is a two-leg spin ladder system composed of 
alternating-spin chains with two kind of 
spins $s_1$ and $s_2$, defined by
%----------------------------------------------------------------------
%   Hamiltonian
%----------------------------------------------------------------------
\begin{equation}
H=\sum_{j=1}^N
\left[
\sum_{i=1}^{2}J\Gamma_{i,j}\bS_{i,j}\cdot\bS_{i,j+1}
+J_0\Gamma_{0,j}\bS_{1,j}\cdot\bS_{2,j}
\right],
\end{equation}
where $\Gamma_{i,j}$ are bond-alternation parameters 
%----------------------------------------------------------------------
%   \Gamma
%----------------------------------------------------------------------
\begin{equation}
\Gamma_{i,j}=\left\{
\begin{array}{ll}
1-\gamma_i & \hbox{for } j= \hbox{odd}\\
1+\gamma_i & \hbox{for } j= \hbox{even}
\end{array}\right. ,
\end{equation}
%%%$\Gamma_{i,j}=1-\gamma_i$ for $j=$odd and $\Gamma_{i,j}=1+\gamma_i$ 
%%%for $j=$even 
for $i=0,1,2$, 
and $N~(=$ even) is the number of sites for each chain.
We denote the spin of the $j$th site 
of the $i$th chain as $s_{i,j}$, where
%----------------------------------------------------------------------
%   s_{i,j}
%----------------------------------------------------------------------
\begin{equation}
s_{i,j}=
\left\{
\begin{array}{ll}
s_1 &  \hbox{for } j= \hbox{odd}\\
s_2 &  \hbox{for } j= \hbox{even}
\end{array}
\right.,
\end{equation}
%%%$s_{i,j}=s_1$ for $j=$odd and $s_{i,j}=s_2$ for $j=$even 
for $i=1,2$.
What is interesting is that the model naturally interpolates 
two kind of alternating-spin chains. 
Namely, when $J_0=0$ it reduces to two independent
alternating-spin chains 
$s_1\otimes s_2\otimes s_1\otimes s_2\otimes\cdots$
with ferrimagnetic ground state\cite{AltThe}.
On the other hand, when 
$\gamma_1=1$ and $\gamma_2=-1$ ($J_0\ne0$) it becomes a single 
``alternating-spin chain'' 
$s_1\otimes s_1\otimes s_2\otimes s_2\otimes\cdots$
with a singlet ground state, which has recently been studied 
in ref.\cite{TonHik,FukKaw1,FukKaw2}.
In what follows, we restrict ourselves to antiferromagnetic couplings
$0<J$, $0<J_0$ and $-1<\gamma_i<1$.
Then the ground state proves to be a unique singlet.

%----------------------------------------------------------------------
%                      III. SPIN-WAVE MODES
%----------------------------------------------------------------------
\section{Spin-wave modes}

Analysis of the spin-waves is indispensable to
mapping to the NLSM, since the mapping
is ensured by the linear-dispersion relation of the
spin-wave mode above the classical N\'eel ground state.  
For this purpose, it is suitable to introduce four
kind of bosons to express the spin generators 
by the use of the Holstein-Primakoff mapping 
(see Fig.\ref{f:SheIll} for the numbering), assuming the N\'eel
configuration.
In the momentum space, the spin-wave Hamiltonian of 
quadratic order in boson operators is calculated as
%----------------------------------------------------------------------
%   spin wave Hamiltonian
%----------------------------------------------------------------------
$
H_{\rm sw}=\sum_{k=1}^{N/2}\bA^\dagger\cH\bA,
$
where 
%----------------------------------------------------------------------
%   \cal H 
%----------------------------------------------------------------------
$
\cH=\left(\begin{array}{cc}h&\Delta\\
\bar{\Delta}&\bar{h}\end{array}\right) .
\label{HamS}%----------------------------------------------------------
$
In this equation, we have defined 
$A_i=a_i~(a_i^\dagger)$ for $i=1\sim4~(5\sim8)$ and  $4\times4$
matrices $h$ and $\Delta$
%----------------------------------------------------------------------
%   h and \Delta
%----------------------------------------------------------------------
\begin{eqnarray}
h=&&J\hbox{diag}(s_2,s_1,s_2,s_1)
\nonumber\\&&\qquad%---------------------------------------------TWOCOL
+\frac{J_0}{2}\hbox{diag}
(\Gamma_{-}s_1,\Gamma_{+}s_2,\Gamma_{-}s_1,\Gamma_{+}s_2),
\nonumber\\
\Delta=&&\frac{1}{2}\left(
\begin{array}{llll}
0&\Delta_{1}&\Delta_{-}&0\\
\Delta_{1}&0&0&\Delta_{+}\\
\Delta_{-}&0&0&\Delta_{2}\\
0&\Delta_{+}&\Delta_{2}&0
\end{array}\right) ,
\end{eqnarray}
where
%----------------------------------------------------------------------
%   \Delta_j and \Delta_\pm 
%----------------------------------------------------------------------
\begin{eqnarray}
&&\Delta_{j}=2J\sqrt{s_1s_2}\left(\cos p+(-)^ji\gamma_j\sin p\right),
\nonumber\\
&&\Delta_{+}=J_0\Gamma_+s_2,
\nonumber\\
&&\Delta_{-}=J_0\Gamma_-s_1.
\end{eqnarray}
We have denoted the momentum as $p\equiv2\pi k/N$ with integers
$k=1,\cdots,N/2$.
Here and in what follows, we occasionally use a simpler notation
$\Gamma_{\pm}\equiv1\pm\gamma_0$.
Now introduce the Bogoliubov transformation
%----------------------------------------------------------------------
%   Bogoliubov transformation
%----------------------------------------------------------------------
$\bA=\cU\bB$, where $\cU$ is a $8\times8$ matrix. 
If the transformed operators $\bB$ satisfy the boson commutation
relations, the matrix $\cU$ should satisfy
%----------------------------------------------------------------------
%   constraint on \cal U
%----------------------------------------------------------------------
$\cU\cN\cU^\dagger=\cU^\dagger\cN\cU=\cN$,
where we have introduced the metric 
$\cN=\hbox{diag}(1_4,-1_4)$ 
with $1_4$ being the $4\times4$ unit matrix.

The spin-waves are composed of four modes in
general. However, in the case
$\gamma_1=-\gamma_2$
two of them become degenerate and we have only two modes.
In what follows, we restrict ourselves to 
this simple case, and investigate especially the 
effects of $\gamma_0$
and its interplay with the spin-alternation for the ground state
properties of the model.
For this purpose, we will simplify the notations,
%----------------------------------------------------------------------
%   restricted bond-alternation
%----------------------------------------------------------------------
\begin{equation}
\gamma_1=-\gamma_2\equiv\gamma_\parallel,\quad
\gamma_0\equiv\gamma.
\end{equation}
For this simple case,
we can diagonalize the spin-wave Hamiltonian
and explicitly obtain the dispersion relations of the form
%----------------------------------------------------------------------
%   explicit spin-wave modes
%----------------------------------------------------------------------
\begin{equation}
\omega_\pm=\sqrt{2}\sqrt{A\sin^2p+B
\pm\sqrt{(2AB-C)\sin^2p+B^2}},
\label{DisRel}%--------------------------------------------------------
\end{equation}
where
%----------------------------------------------------------------------
%   parameters
%----------------------------------------------------------------------
\begin{eqnarray}
A=&&2J^2(1-\gamma_\parallel^2)s_1s_2\nonumber\\
B=&&J[J(s_1-s_2)^2+2J_0s_1s_2]\nonumber\\
C=&&4J^2J_0s_1s_2
\{J(1-\gamma_\parallel^2)
[(1-\gamma)s_1^2+(1+\gamma)s_2^2]\nonumber\\
&&+J_0(1-\gamma^2)s_1s_2\}
\end{eqnarray}
The lower spin-wave mode in eq.(\ref{DisRel}) is thus found 
to have a linear dispersion for small momenta, $\omega_-\sim v_sp$, 
where 
\begin{equation}
v_s=\sqrt{C/B}.
\label{Velo}%--------------------------------------------------------
\end{equation}
%--------------------------------------------------------------------
%   FIGURE of spin wave mode WITHOUT bond-alternation
%--------------------------------------------------------------------
\begin{figure}[htb]
\epsfxsize=6cm %%% 6 is suitable for 2 column
\centerline{\epsfbox{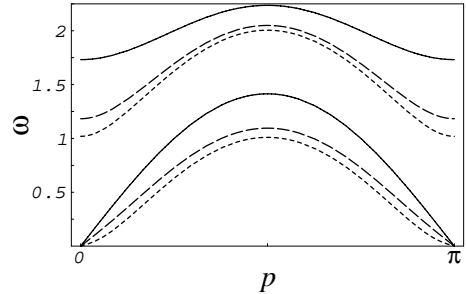}} 
\vspace{0.2cm}
\caption{Spin-wave spectrum  
as functions of the momentum $p$ for
$J_0=0.01$ (dotted line)
0.1 (dashed line) and 0.5 (solid line)
with other parameters being fixed as $s_1=1/2$, $s_2=1$, $J=1$ 
and $\gamma_0=\gamma_1=\gamma_2=0$.}
\label{f:Mod}%-------------------------------------------------------
\end{figure}
In Fig.\ref{f:Mod}, we present some examples of these modes. 
First, let us look at the dotted line in Fig.\ref{f:Mod}, 
which is calculated for rather small inter-chain coupling $J_0=J/100$. 
Note that it resembles, in this energy scale, the 
quadratic dispersion relation for the ferrimagnetic case,
%%which is given by 
%----------------------------------------------------------------------
%   dispersion of alternating-spin chain 
%----------------------------------------------------------------------
$
\omega_\pm/J=\pm|s_2-s_1|+
\left[(s_1-s_2)^2+4s_1s_2\sin^2p\right]^{1/2}
$.
However, from eq.(\ref{DisRel}), the lower
mode is always linear as far as $J_0\ne0$, even though its linearity is
restricted to small momentum region.
This reflects the fact that the ground state is always
singlet in spite of the magnitude of $J_0(\neq 0)$. In fact,
if we increase the inter-chain coupling $J_0$ up to $\sim J/10$
(dashed-line), we can explicitly see that 
the lower mode has a linear dispersion in a wider
momentum region.
Therefore, even for rather small $J_0$, we expect that the model
can be described by the NLSM at least qualitatively.

%-----------------------------------------------------------------------
%                    III. NONLINEAR SIGMA MODEL
%-----------------------------------------------------------------------
\section{O(3) nonlinear sigma model approach}

We have so far investigated the classical properties of the
ground state. 
However, the low-energy quantum fluctuation
plays an essential role in determining the true ground state.
In order to investigate them, we will use 
NLSM techniques\cite{Hal,Rev,Sene,DSS,DEMPR}.
Introducing the SU(2) coherent state by 
$\langle\bn_{i,j}|\bS_{i,j}|\bn_{i,j}\rangle
=s_{i,j}(-)^{i+j}\bn_{i,j}$, 
we have the effective action given by $S=S_B+S_I$, where
%----------------------------------------------------------------------
%   action
%----------------------------------------------------------------------
\begin{eqnarray}
&&S_B=-i\sum_{i=1}^2\sum_{j=1}^Ns_{i,j}(-)^{i+j}\omega[\bn_{i,j}],
\nonumber\\
&&S_I=-\sum_{j=1}^N\biggl(\sum_{i=1}^2J\Gamma_{i,j}\int_0^\beta\!\! 
d\tau\bn_{i,j}\scdot\bn_{i,j+1}
\nonumber\\&&\qquad\qquad\qquad\qquad%---------------------------TWOCOL
+J_0\Gamma_{0,j}\int_0^\beta\!\! 
d\tau\bn_{1,j}\scdot\bn_{2,j}\biggr),
\end{eqnarray}
Here, $\omega[\bn]$ is the Berry phase defined by
%----------------------------------------------------------------------
%   Berry phase
%----------------------------------------------------------------------
$
%\begin{equation}
\omega[\bn]=\int_0^\beta\!\!d\tau\int_0^1\!\!du
\bn\scdot(\partial_\tau\bn\stimes\partial_u\bn).
%\end{equation}
$

Now introduce the semi-classical N\'eel configuration $\bm$
and fluctuations $\bl_a$ around it as follows,
%----------------------------------------------------------------------
%   m and l fields
%----------------------------------------------------------------------
\begin{equation}
\bn_{i,2j+a}=\bm(2j+a)+(-)^{i+a}a_0\bl_a(2j+a),
\end{equation}
for $i=1,2$, $j=0,1,\cdots,N/2-1$ and for $a=1,2$, where
$a_0$ is a lattice constant.
We have introduced two kind of the fluctuation fields $\bl_1$ and
$\bl_2$\cite{Sene,DSS,DEMPR}, by taking into account 
the fact that we have two spin-wave modes
in the above spin-wave analysis.
Now the calculation is straightforward, and we reach the lagrangian
%----------------------------------------------------------------------
%   Lagrangian
%----------------------------------------------------------------------
\begin{eqnarray}
{\cal L}=&&\
\frac{1}{2}\bl_aL_{ab}\bl_b
+iu_a\bl_a\scdot(\bm\stimes\partial_\tau\bm)
\nonumber\\
&&+v_a\bl_a\scdot\partial_x\bm
+w(\partial_x\bm)^2
\end{eqnarray}
%----------------------------------------------------------------------
%   parameters in Lagrangian
%----------------------------------------------------------------------
where repeated indices $a$ and $b$ should be summed over, and where
\begin{eqnarray}
&&L=2\left(\begin{array}{cc}
Js_1s_2+J_0\Gamma_{-}s_1^2&Js_1s_2\\
Js_1s_2&Js_1s_2+J_0\Gamma_{+}s_2^2
\end{array}\right),
\nonumber\\
&&\bu^t=(s_1,s_2),\quad
\bv^t=2J\gamma_\parallel s_1s_2(1,1),\quad
w=Js_1s_2.
\end{eqnarray}
Here, the couplings $J$ and $J_0$ mean those
in unit of $a_0$.

Integrating out the fields $\bl$, we end up with 
%----------------------------------------------------------------------
%   final lagrangian
%----------------------------------------------------------------------
\begin{eqnarray}
{\cal L}=&&\frac{1}{2g}\left[
v_s(\partial_1\bm)^2+\frac{1}{v_s}(\partial_2\bm)^2\right]
\nonumber\\&&\qquad\qquad%---------------------------------------TWOCOL
+\frac{\theta}{8\pi}\epsilon_{\mu\nu}\bm
\cdot(\partial_\mu\bm\stimes\partial_\nu\bm),
\label{Lag}%-----------------------------------------------------------
\end{eqnarray}
where
%----------------------------------------------------------------------
%   formula of \theta, g and v
%----------------------------------------------------------------------
\begin{eqnarray}
\theta&=&4\pi i\bu^tL^{-1}\bv
\nonumber\\
&=&
\frac{4\pi i\gamma_\parallel s_1s_2
\left[(1-\gamma)s_1+(1+\gamma)s_2\right]}
{(1-\gamma)s_1^2+(1+\gamma)s_2^2+R(1-\gamma^2)s_1s_2},
\nonumber\\
g&=&\left[(2w-\bv^tL^{-1}\bv)\bu^tL^{-1}\bu\right]^{-1/2},
\nonumber\\
v_s&=&\left[(2w-\bv^tL^{-1}\bv)/\bu^tL^{-1}\bu\right]^{1/2}.
\label{TGV}%-----------------------------------------------------------
\end{eqnarray}
Here we have defined the ratio  $R=J_0/J$.
What is remarkable is that the spin-wave velocity calculated here 
coincides exactly with eq.(\ref{Velo}),
which implies that the present NLSM approach 
is consistent with the spin-wave analysis.
Before discussing the phase diagram, we will 
check the formulae (\ref{TGV}) 
by comparing them with those in some limits we have already known.
First, set $s_1=s_2\equiv s$ and $\gamma=0$, and we have 
$\theta/\pi i=8s\gamma_\parallel/(2+R)$, 
exactly coinciding with the formula in
ref.\cite{DSS} for the usual two-leg ladder with intra-chain
bond-alternation. 
Set furthermore $\gamma_\parallel=0$, we have 
$\theta=0$, $g=1/s\times(1+R/2)^{1/2}$ and $v_s=2sJ(1+R/2)^{1/2}$, 
which correctly reproduce the formulae in \cite{Sene,DEMPR} 
for the usual uniform ladder.
Next, set $\gamma_\parallel=1$, $\gamma=0$, 
$J=J'(1+\gamma')/2$, and $J_0=J'(1-\gamma')$, 
then we have the expression
$\theta=4\pi i(1+\gamma')s_1s_2(s_1+s_2)/[(s_1+s_2)^2+\gamma'
(s_1-s_2)^2]$, etc, 
derived in \cite{FukKaw2} for the single 
$s_1\otimes s_1\otimes s_2\otimes s_2\otimes\cdots$
chain.\cite{Com1}

%---------------------------------------------------------------------
%                   IV. PHASE DIAGRAM
%---------------------------------------------------------------------
\section{Phase diagram}

Now let us investigate various phases of the model.
First of all, the formula (\ref{TGV}) tells us that if
$\gamma_\parallel=0$ we have always $\theta=0$,
leading to massive phases.
In other words, {\it non-trivial massless phases can occur if
$\gamma_\parallel\ne0$.}
In what follows, we assume $\gamma_\parallel\ne0$
and investigate the phases of the model.
Let us first consider the case $s_2\rightarrow\infty$, 
for example. We have 
$\theta\rightarrow4\pi i\gamma_\parallel s_1$, i.e,
the phases of the model are controlled 
only by the smaller spin, independent of $R$ and $\gamma$.
This simple statement is, needless to say, 
valid only for this case:
In the rest of the paper, we investigate the properties of the
phase transitions in more general cases, putting stress on how the
spin-alternation affects them.
In what follows, we set $s_1\le s_2$ for simplicity.
As the intra-chain bond-alternation $\gamma_\parallel$
has been already discussed in ref.\cite{DSS}, 
we concentrate on the inter-chain bond-alternation $\gamma~(=\gamma_0)$ 
and its interplay with the spin-alternation, assuming
$0\le\gamma_\parallel$ without loss of generality.  
Setting $\theta/(\pi i)\equiv n$ with $n$ being an odd integer,
we can draw in the $R$-$\gamma$ plane the critical lines on which the
model is expected to be massless,
%----------------------------------------------------------------------
%   massless R-\gamma line
%----------------------------------------------------------------------
\begin{equation}
R=\frac{s_1}{1+\gamma}\left(\frac{4\gamma_\parallel}{n}
-\frac{1}{s_2}\right)
+\frac{s_2}{1-\gamma}\left(\frac{4\gamma_\parallel}{n}
-\frac{1}{s_1}\right).
\end{equation}
We can see that according to the value of $n$
there appear three kind of lines (see Fig.\ref{f:Phdia2}), 
\begin{enumerate}
\item[(i)] $n<4\gamma_\parallel s_1$,
\item[(ii)] $n=4\gamma_\parallel s_1$, 
\item[(iii)] $4\gamma_\parallel s_1<n<4\gamma_\parallel s_2$.
\end{enumerate}
The line (ii) exists only the case in which $4\gamma_\parallel s_1$
happens to be an odd integer.  The lines (ii) and (iii) 
can exist only if the two
kind of the spins are different, $s_1\ne s_2$.
In other words, {\it 
the effects of the spin-alternation can be observed  
typically by the existence of lines of the types} (ii) {\it and} (iii).

We note that in the region near the $\gamma$ axis
(with quite small $R$), 
the present approximation may become worse, because
when exactly $R=0$, the model decouples to two independent 
alternating-spin chains with ferrimagnetic ground state, 
which cannot be described by the present approach.
However, as noted above, as long as $J_0\ne0$ we always have a linear
spin-wave dispersion even though it is
restricted, for small $J_0$, in a small momentum region.
Therefore, we believe 
that the phase diagram in the $R$-$\gamma$ plane is 
qualitatively valid even for small $R$.

%---------------------------------------------------------------------
%                     s_1=s_2
%---------------------------------------------------------------------
\subsection{$s_1=s_2$ case}

Because the inter-chain bond-alternation is introduced,
as far as we know, 
for the first time in this paper,
let us first consider the usual ladder system with $s_1=s_2$. 
What we would like to claim here is that
the inter-chain bond-alternation really affects $\theta$ once 
the finite intra-chain bond-alternation is introduced
even for the usual $s_1=s_2=1/2$ ladder as follows:
\begin{enumerate}
\item[(1)] for $1/2<\gamma_\parallel<1$ we have only one massless line
$R=2(2\gamma_\parallel-1)/(1-\gamma^2)\equiv R_1(\gamma)$, 
corresponding to $n=1$, 
\item[(2)] for $0\le\gamma_\parallel\le1/2$ we have no massless line.
\end{enumerate}
In case (1), we expect to have 
the VBS state 
%----------------------------------------------------------------------
%   VBS 1
%----------------------------------------------------------------------
\vspace{1mm}\\
\centerline{\epsfbox{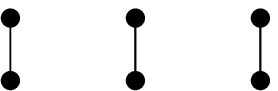}}
in the region $R>R_1(\gamma)$
while in the opposite region $R<R_1(\gamma)$ we may have the state
%----------------------------------------------------------------------
%   VBS 2
%----------------------------------------------------------------------
\vspace{1mm}\\
\centerline{\epsfbox{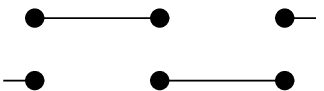}}
between which we have a massless phase.
Contrary to this, in case (2), we have the former VBS state in the
whole $R$-$\gamma$ plane. From these observations, we see that the
inter-chain bond-alternation indeed affects the phase structure.

%----------------------------------------------------------------------
%        s_1\ne s_2
%----------------------------------------------------------------------
\subsection{$s_1\ne s_2$ case}

Now let us discuss the model with $s_1=1/2$ and $s_2=1$, which may be
one of the most probable candidates for the experimental observation
of the alternating-spin ladders. 
%----------------------------------------------------------------------
%   FIGURE of phase diagram: s1=1/2, s2=1.
%----------------------------------------------------------------------
\begin{figure}[htb]
\epsfxsize=6.0cm %%% 6 is suitable for 2 column
\centerline{\epsfbox{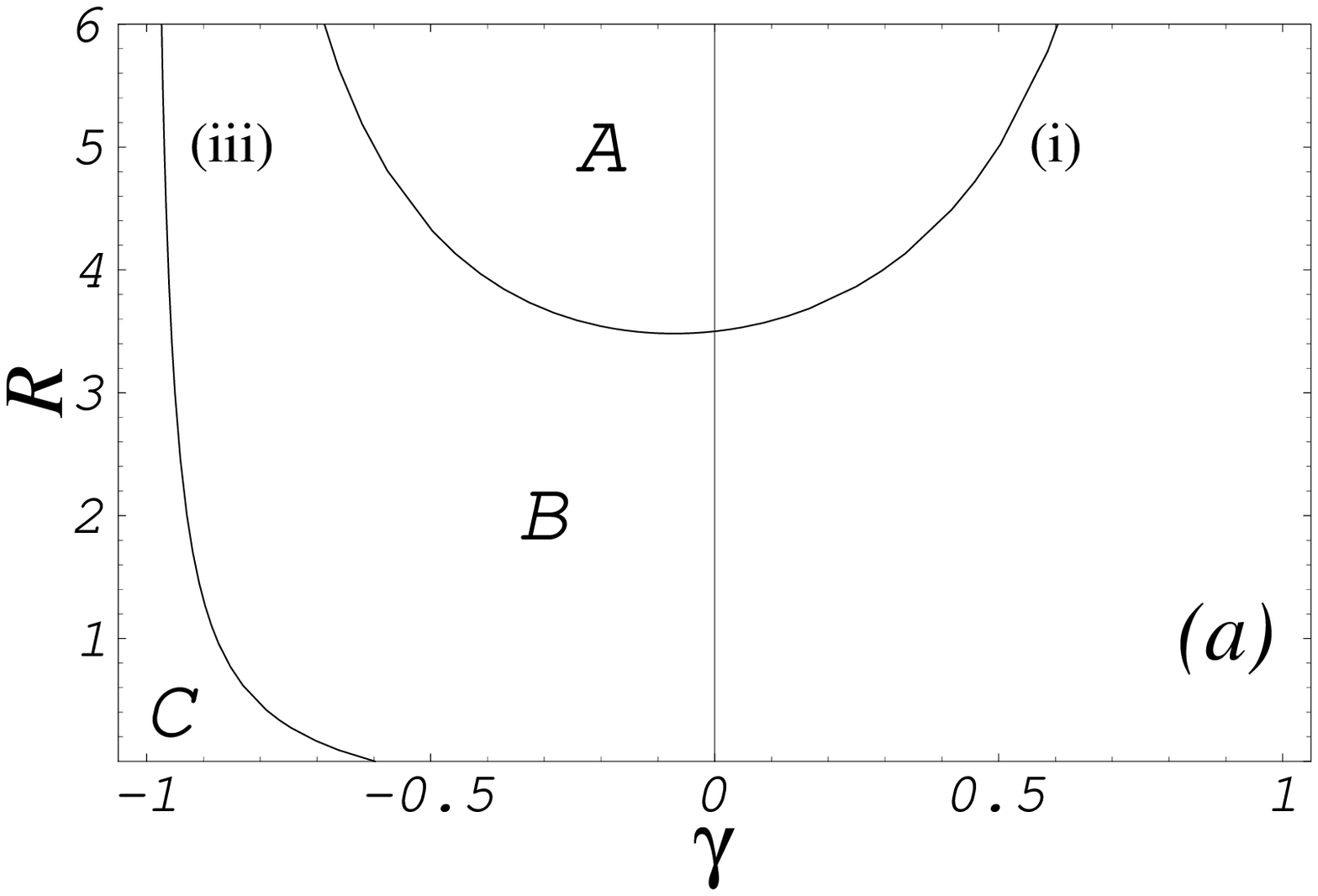}}
\epsfxsize=6.0cm %%% 6 is suitable for 2 column
\centerline{\epsfbox{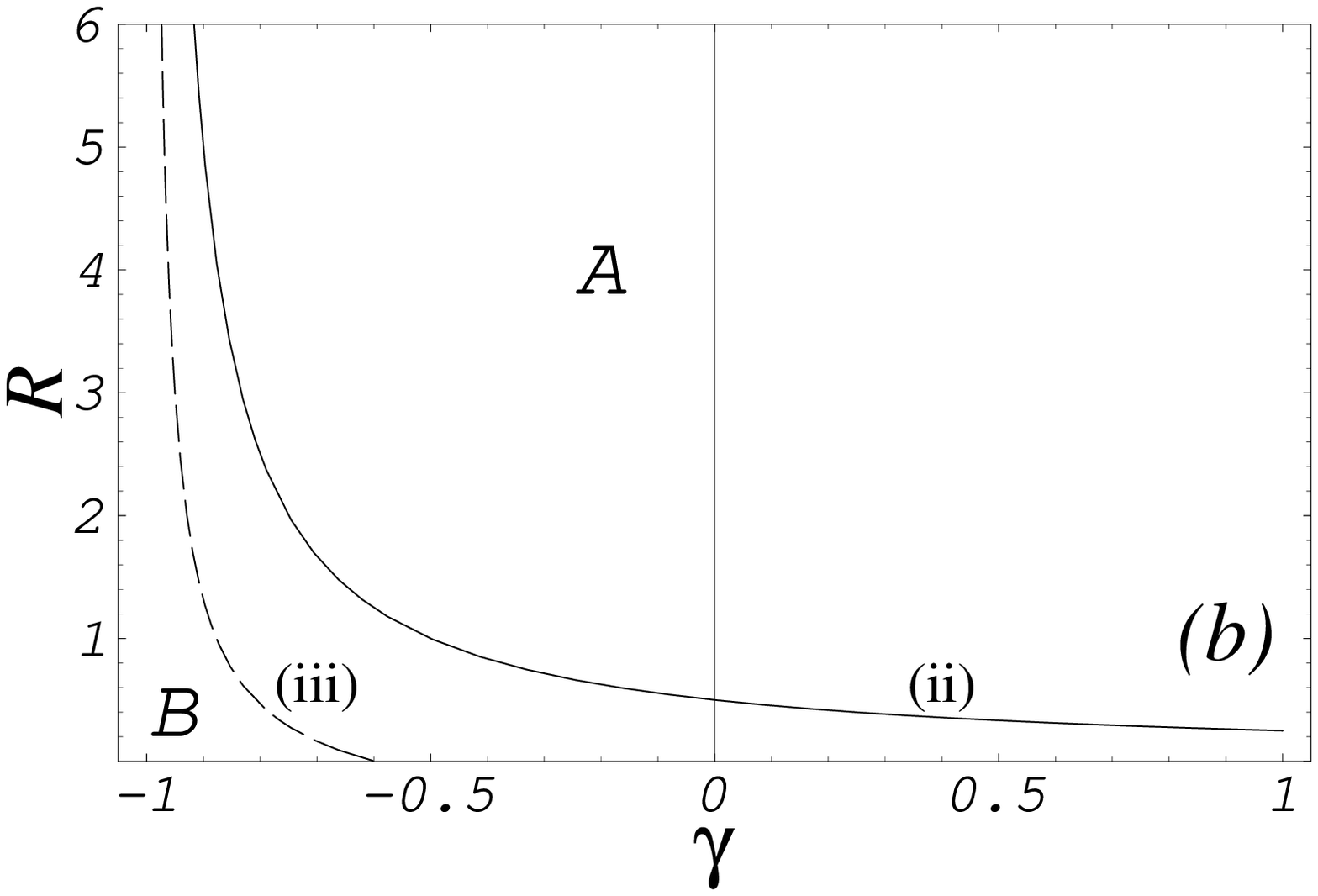}}
\vspace{0.2cm}
\caption{
The phase diagram for $s_1=1/2$ and $s_2=1$.
(a): $\gamma_\parallel=1$. 
(b): $\gamma_\parallel=1/2$ for solid
line and $\gamma_\parallel=1/3$ for dashed line.
}
\label{f:Phdia2}%------------------------------------------------------
\end{figure}
The phase diagram is qualitatively classified into 
(I) $3/4<\gamma_\parallel\le1$, 
(II) $1/2<\gamma_\parallel\le3/4$,
(III) $\gamma_\parallel=1/2$,
(IV) $1/4<\gamma_\parallel<1/2$ and
(V) $\gamma_\parallel\le1/4$,
according to how many critical lines of the categolies (i), (ii) and
(iii) appear. 
In Fig.\ref{f:Phdia2}, we present some examples of the phase diagram.
In Fig.\ref{f:Phdia2}(a), there appear three kind of massive phases 
separated by two massless lines (case (I)). 
If we trace the phase diagram along the $R$-axis ($\gamma=0$ line), 
we can easily specify the phases A and B 
%----------------------------------------------------------------------
%   figure of VBS 
%----------------------------------------------------------------------
\begin{figure}[htb]
\begin{center}
\epsfxsize=6.0cm %%% 6 is suitable for 2 column
\centerline{\epsfbox{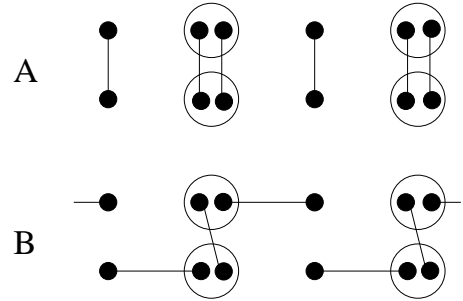}}
\end{center}
\caption{VBS states for the phases A and B.}
\label{f:VBSAB}%------------------------------------------------------
\end{figure}
in terms of VBS picture in Fig.\ref{f:VBSAB}
(in the case $\gamma_\parallel=1$ see \cite{TonHik}).
What is most interesting is the appearance of the 
new phase C, separated from the others
by another massless line which belongs to the category (iii).
Namely this  phase is caused by the interplay between
the bond-alternation and  the spin-alternation.
In order to see what kind of the ground state is 
indeed realized in the phase C, 
let us first start with a point in the 
region A, and move parallel to $\gamma$-axis 
toward the $\gamma=-1$ direction.  In this process,
the interaction between two
$s_2=1$ spins is decreased, so that 
one of the valence-bond changes its partner, and we thus
reach the phase B. Then the question is what happens 
when the interaction between two $s_1=1/2$ spins
is further increased. 
We then expect that $s_1=1/2$ spins form the valence-bond,
and as a consequence $s_2=1$ spins cannot help forming the 
valence-bond again, which results in  the phase C. Therefore, 
it is natural to conclude
that the resulting ground state in the 
phase C is identical to that in the phase A.
So far we have discussed the phase diagram of (I). The diagram of 
(II) is similar to (I), but without the line (iii) 
in Fig.\ref{f:Phdia2}(a) and hence without the phase C.

Shown in Fig.\ref{f:Phdia2}(b) is another example of the
phase diagram for which the critical line is essentially 
determined by  the
spin-alternation itself (cases (III) and (IV)). 
What is to be noted is the difference between 
two cases, the  critical line for which is indicated by 
the solid line ($\gamma_\parallel=1/2$) 
and the dashed line $(\gamma_\parallel=1/3)$. 
For example, if we change the ratio of 
$R$ with $\gamma=0$ being fixed, we encounter the
phase transition in the former case, 
while in the latter case we are always in the A phase.
Also, if we start from $\gamma=-1$ with rather small
 $R$ towards $\gamma=1$ direction, 
we may be always in the phase B 
in the former case, while in the latter case we 
experience the transition from the phase B to
A before $\gamma=0$.
We note that
the phase diagram of (V) has no critical lines, and hence has 
only the phase A.

%---------------------------------------------------------------------
%                   V. DISCUSSIONS
%---------------------------------------------------------------------
\section{Summary and discussions}

We have proposed a spin-alternating ladder model, which shows
interesting interplay between 
the bond-alternation and the spin-alternation.
We have indeed shown how the spin-alternation affects the quantum
phase transitions, and have derived the phase diagram by means of 
O(3) nonlinear sigma model techniques.

Here, some comments are in order. First we wish to mention the effects 
of frustration which have been ignored in this paper.
For example, if we include the next-nearest interaction,
some interesting phenomena may be expected to happen. 
However, as far as such frustration is small enough to 
be treated as a perturbation, it merely renormalizes 
the coupling $g$ and the velocity $v_s$, for which the present 
conclusion may be still valid. The detailed study on the 
effect of frustration is an interesting issue to be
explored in the future study. 
Another comment is on the quantitative arguments for the phase
diagram. Even for ordinary spin chains, the predicted 
values by the NLSM for the bond-alternation parameter 
which causes massless phases\cite{AffHal} are slightly different from 
those obtained by numerical calculations\cite{KatTan}.
Nevertheless, it is believed that the NLSM can describe  
qualitatively correct low-energy properties, and 
the number of the massless phases 
in varying the bond-alternation parameter is correctly
predicted. We believe that these statements are also the case for
our model, and the present analysis based on the NLSM 
should provide  the qualitatively correct
phase diagram, although the predicted critical lines
should be somehow modified by more accurate treatment.

To conclude the paper, we wish to
emphasize again that although the model we proposed in this paper 
seems somewhat complicated at first sight,
it is a generalized model which
naturally interpolates various interesting quantum spin systems
investigated intensively. 
For example, if we set $\gamma=1$, the model is reduced to a single
chain composed of two kind of spins, while for the 
case $\gamma=0$ and $s_1=s_2$ it becomes a usual ladder model.
What is the merit of studying the present model is that a 
wider class of spin models can be treated on an equal footing.
So far, spin ladder systems with the spin-alternation 
have  not been found experimentally.  We hope 
that a ladder system proposed 
here, or a system which naturally interpolates 
spin chains  and ladders can be realized experimentally 
in the near future. These issues then 
provide new paradigms of the quantum phase transitions 
for low-dimensional spin systems.
In such cases,  the present model could
serve as a key model which connects the physics of 
various spin systems such as the spin ladder, the alternating 
spin chain, etc.

%----------------------------------------------------------------------
%                   Acknowledgments
%----------------------------------------------------------------------
\acknowledgements
The authors would like to thank T. Tonegawa, M. Kaburagi and 
M. Chiba for valuable discussions.
This work is partly
supported by the Grant-in-Aid from the Ministry of
Education, Science and Culture, Japan.

%----------------------------------------------------------------------
%   FIGURES
%----------------------------------------------------------------------


\begin{references}
\bibitem{Com1} Note that 
the present formula for $v_s$ is half as large as 
that in \cite{FukKaw2}, because of
the difference of the definition of the momentum. 
In the snake chain ($\gamma_\parallel=1$),
the length of the system is
twice as long as the usual single chain.
\end{references}
\end{document}